\documentclass[sigconf]{acmart}

\renewcommand\footnotetextcopyrightpermission[1]{}
\acmConference[Preprint]{Preprint}{April}{2026}

\AtBeginDocument{%
  }

\acmISBN{978-1-4503-XXXX-X/2026/06}

\begin{document}

\title{Quality-Driven Selective Mutation for Deep Learning}
\author{Zaheed Ahmed}
\orcid{0000-0001-6594-537X}
\affiliation{%
  \institution{Institute of Computer Science, University of Göttingen}
  \city{Goettingen}
  \state{Lower Saxony}
  \country{Germany}}
\email{zaheed.ahmed@informatik.uni-goettingen.de}

\author{Emmanuel Charleson Dapaah}
\orcid{0009-0005-5374-311X}
\affiliation{%
  \institution{Institute of Computer Science, University of Göttingen}
  \city{Goettingen}
  \state{Lower Saxony}
  \country{Germany}}
\email{dapaah@cs.uni-goettingen.de}

\author{Philip Makedonski}
\orcid{0000-0001-7752-0029}
\affiliation{%
  \institution{Institute of Computer Science, University of Göttingen}
  \city{Goettingen}
  \state{Lower Saxony}
  \country{Germany}}
\email{makedonski@informatik.uni-goettingen.de}

\author{Jens Grabowski}
\orcid{0000-0003-2994-3531}
\affiliation{%
  \institution{Institute of Computer Science, University of Göttingen}
  \city{Goettingen}
  \state{Lower Saxony}
  \country{Germany}}
\email{grabowski@informatik.uni-goettingen.de}

\renewcommand{\shortauthors}{Ahmed et al.}

\begin{abstract}
Mutants support testing and debugging in two roles: (i) as test goals and (ii) as substitutes for real faults.
Hard-to-kill mutants provide better guidance for test improvement, while realism is essential when mutants are used to simulate real bugs.
Building on these roles, selective mutation for deep learning (DL) aims to reduce the cost of mutant generation and execution by choosing operator configurations that yield resistant and realistic mutants.
However, the DL literature lacks a unified measure that captures both aspects.

This study presents a probabilistic framework to quantify mutant quality along two complementary axes: \emph{resistance} and \emph{realism}.
Resistance adapts the classical notion of hard-to-kill mutants to the DL setting using statistical killing probabilities, while realism is measured via the generalized Jaccard similarity between mutant and real-fault detectability patterns.
The framework enables ranking and filtering of low-quality mutation-operator configurations without assuming a specific use case.

We empirically evaluate the approach on four datasets of real DL faults.
Three datasets (\texttt{CleanML}, \texttt{DeepFD}, and \texttt{DeepLocalize}) are used to estimate and select high-quality operator configurations, and the held-out \texttt{defect4ML} dataset is used for validation.
Results show that quality-driven selection reduces the number of generated mutants by up to~55.6\% while preserving typical levels of resistance and realism under baseline-aligned selection thresholds.
These findings confirm that dual-objective selection can lower cost without compromising the usefulness of mutants for either role.
\end{abstract}



\maketitle

\section{Introduction}
\label{sec:introduction}
Mutation testing implements the concept of introducing artificial bugs to support testing and debugging activities. In this approach, synthetic faults are injected into the system under test using mutation operators to create faulty program versions known as \textit{mutants}. 
Typically, mutants serve two main purposes: they act as \emph{test goals} to evaluate and improve test suites and as \emph{substitutes for real faults} to support experimentation and empirical evaluation \citep{jia_analysis_2011, papadakis_mutation_2019}.
Based on their role, selection of mutants is different in each case. Prior studies indicate that, for using mutants as \emph{test goals}, prioritizing mutants based on their degree of resistance yields better guidance for test improvement \cite{kaufman_prioritizing_2022, petrovic_practical_2022, kurtz_analyzing_2016, estero-botaro_quality_2015}. 
For using mutants as \emph{substitutes} in evaluation of techniques (e.g., fault localization, program repair), \emph{realism} is essential for valid conclusions \cite{brown_care_2017, allamanis_tailored_2016, just_are_2014}. The computational cost incurred in executing each mutant against the test suite is a major drawback. While mutation testing has matured for conventional software, adapting it to data-driven systems raises new scalability and realism challenges.

Recently, mutation analysis has been adapted for \emph{deep learning (DL)} systems \citep{ma_deepmutation_2018, shen_munn_2018, hu_deepmutation_2019, humbatova_deepcrime_2021, kim_muff_2025}. Beyond their traditional role as test goals, DL mutants have been employed to evaluate a diverse range of techniques, including automated program repair \citep{wu_mutation-based_2021, sohn_arachne_2023}, input data prioritization for labeling \citep{wang_prioritizing_2021}, robustness evaluation of neural models \citep{hu_deepmutation_2019, lin_robustness_2022, mendez_testing_2024}, 
assessment of data cleaning techniques \citep{li_cleanml_2021, abdelaal_rein_2023}, generation and detection of adversarial examples \citep{wang_adversarial_2019, pour_search-based_2021}, test data generation \citep{riccio_deepmetis_2021, deokuliar_improving_2023}, differential testing of DL libraries \citep{wang_deep_2020}, structural analysis of deep neural networks \citep{ghanbari_decomposition_2024}, oracle generation for autonomous vehicles \citep{jahangirova_quality_2021}, fault localization in deep neural networks \citep{cao_deepfd_2022, ghanbari_mutation-based_2023}, and support for metamorphic testing strategies in machine learning \citep{xie_testing_2011}.
Collectively, these diverse uses of mutants emphasize the need for efficiently generating high-quality and useful mutants.

In DL, mutants can be generated using either \emph{pre-training} or \emph{post-training} approaches \citep{ma_deepmutation_2018, ahmed_exploring_2024}. Empirical evidence shows that pre-training mutants are generally more realistic than their post-training counterparts~\citep{ahmed_empirical_2025}. However, computational cost is a more severe issue in DL than in classical software systems. While executing mutants is already expensive, generating mutants with pre-training operators is even more computationally demanding. Moreover, due to the statistical killing criterion~\cite{jahangirova_empirical_2020} employed in the pre-training mutation testing tool DeepCrime~\cite{humbatova_deepcrime_2023}, each model must be trained multiple times, further adding to the cost. For instance, generating and analyzing pre-training mutants with DeepCrime on a single LeNet-5 model trained on MNIST required more than 123~hours, even with a limited input set, highlighting the substantial computational overhead of pre-training mutation analysis~\cite{abbasishahkoo_teasma_2024}. Similarly, Shen et~al.~\cite{shen_boundary_2021} applied DeepMutation~\cite{ma_deepmutation_2018}, which incorporates both pre- and post-training mutation operators, to three DL models. Their experiment generated nearly 5{,}000 mutants, each evaluated on 10{,}000 test samples, totaling about 50~million test executions and taking roughly three weeks to complete on a high-performance server. This motivates the need for cost-effective strategies such as selective mutation.

Reducing the number of generated mutants is essential for making mutation testing of DL systems scalable. 
In conventional software testing, \emph{selective mutation} mitigates cost by excluding less effective mutation operators while preserving test adequacy~\citep{jia_analysis_2011, papadakis_mutation_2019}.
Following the same rationale, DL mutation testing can benefit from a careful analysis of mutation-operator configurations. Pre-training mutation operators are often parametric, and each configuration injects a distinct fault.
Since not all configurations are equally useful, killed mutants also vary in their informativeness: some are trivial and easily killed, whereas others require more specific and sensitive tests. Extending selective mutation to the DL domain enables us to retain only the most useful operator configurations, thereby reducing the number of mutants that must be generated and executed. This reduction alleviates the computational burden and makes mutation testing more practical for large-scale DL systems.

Building on the use cases of mutants, the quality of DL mutants should be characterized along two complementary dimensions:
(i)~\textit{intrinsic quality (IQ)}, reflecting a mutant’s \emph{resistance} to killing, and
(ii)~\textit{extrinsic quality (EQ)}, reflecting its \emph{realism} relative to real faults.
A desirable subset of mutation-operator configurations should therefore yield mutants that are both hard-to-kill and realistic.
However, existing work provides no unified framework for quantifying these two aspects jointly.

In this study, we present a unified probabilistic framework to measure \emph{IQ} and \emph{EQ}, enabling us to filter out low-quality configurations of mutation operators.
On one hand, we adapt the classical notion of the \emph{resistant mutant} from traditional mutation testing~\citep{estero-botaro_quality_2015}, which defines mutant quality in terms of how difficult a mutant is to kill and how discriminating the killing tests are.
We extend this notion and its deterministic DL adaptation~\citep{zhang_fine-grained_2025} into a probabilistic formulation that captures the inherently statistical nature of mutant killing in DL systems.
On the other hand, we use the generalized Jaccard similarity~\citep{castro_fernandez_lazo_2019} to quantify the behavioral realism of mutants with respect to real faults.
Combined, these two measures enable us to prioritize the configurations of mutation operators. 
The key contributions of this study are as follows:

\begin{itemize}
    \item A unified probabilistic framework for quality-driven selective mutation in DL, jointly characterizing mutant resistance and behavioral realism.
    \item An adaptation of the classical resistant mutant notion to the statistical killing regime of DL systems.
    \item An empirical evaluation on the \textit{defect4ML} dataset demonstrating that our dual-axis quality measures reduce the number of generated mutants while maintaining their utility for both testing and debugging use cases.
\end{itemize}

The rest of this paper is organized as follows.
Section~\ref{sec:related_work} reviews the related work, followed by
Section~\ref{sec:methodology}, which describes the proposed methodology.
Section~\ref{sec:experiment_design} then presents the experimental design,
and Section~\ref{sec:results_analysis} reports and analyzes the empirical results.
Section~\ref{sec:discussion} discusses the implications and limitations of the findings,
Section~\ref{sec:threats} outlines threats to validity,
and Section~\ref{sec:conclusion} concludes the paper.

\section{Related Work}
\label{sec:related_work}
This work relates most closely to research on \textit{quality-based selective mutation} for DL systems, which aims to reduce computational cost while preserving the usefulness of generated mutants. Accordingly, prior related work can be grouped into two main directions: (i) approaches that seek to improve scalability and cost-efficiency, and (ii) studies that attempt to quantify the quality of DL mutants.

\subsection{Cost Reduction in DL Mutation Testing}
Frameworks such as DeepMutation~\citep{ma_deepmutation_2018}, MuNN~\citep{shen_munn_2018}, DeepMutation++~\citep{hu_deepmutation_2019}, DeepCrime~\citep{humbatova_deepcrime_2021}, and MUFF~\citep{kim_muff_2025} demonstrated the potential of mutation analysis for DL models but also revealed its high computational overhead. 
Pre-training mutation approaches require retraining each mutant from scratch, whereas post-training approaches still demand large-scale test execution for every mutated model. 
These costs have motivated several studies to make DL mutation testing more efficient and scalable.

\citet{shen_boundary_2021} proposed \emph{boundary sampling}, selecting only test inputs close to the decision boundary to reduce execution effort without compromising detection effectiveness. 
Similarly, \citet{li_how_2022} investigated higher-order mutation testing for DL systems, showing that certain second-order mutants can represent multiple first-order mutants and thus reduce the number of executions required. 
Likewise, \citet{feng_mutation_2022} analyzed decision-boundary changes to identify operators that cause minimal behavioral variation and can be safely excluded. 
\citet{wang_fine-grained_2023} conducted a fine-grained evaluation of DL mutation operators, introducing redundancy and quality scores to guide operator selection. 
\citet{zhang_fine-grained_2025} extended this idea by quantifying operator quality through aggregated per-mutant scores and using them to prune less effective operators, achieving cost reduction via quality-based selective mutation. 
\citet{xue_deepweak_2024} proposed \textit{DeepWeak}, which reduces execution time by adopting weak mutation testing for DL models, detecting behavioral deviations earlier in the execution pipeline. 
More recently, \citet{lyons_accelerating_2025} accelerated mutation analysis by clustering neurons and mutants with similar behavior, reducing execution time without loss of coverage.

Taken together, these studies enhance scalability by limiting the number of test inputs or mutants to execute. 
Early approaches~\citep{shen_boundary_2021, li_how_2022, feng_mutation_2022} rely primarily on structural and rule-based criteria to reduce computational cost. 
Later work~\citep{wang_fine-grained_2023, zhang_fine-grained_2025} introduced quantitative criteria for selective mutation, but their formulations remain focused at the operator level and are deterministic. 
They do not capture the stochastic behavior of DL training or differentiate between a mutant’s intrinsic resistance and extrinsic realism. 
This work advances the field by linking cost reduction to a probabilistic assessment of mutant quality, enabling configuration-level selective mutation guided by both resistance and realism.

\subsection{Quality Measurement of DL Mutants}

Beyond cost efficiency, recent research has focused on assessing the \emph{quality} of DL mutants to determine which ones are most informative. 
\citet{jahangirova_empirical_2020} conducted one of the earliest empirical analyses of DL mutation operators, showing that operator \emph{type} and \emph{configuration} significantly affect mutant behavior. 
Their study identified operator configurations that produce meaningful, non-trivial behavioral deviations but also revealed that the single-run notion of killing used in earlier work~\citep{ma_deepmutation_2018} is unreliable due to the stochastic nature of DL training. 
They therefore advocated multi-run evaluation to obtain statistically stable and comparable results.

\citet{humbatova_deepcrime_2021} advanced this direction with \textit{DeepCrime}, introducing a probabilistic definition of mutant killing aggregated across multiple retrainings. 
They quantified mutant \emph{killability}, \emph{triviality}, and \emph{redundancy} under this statistical-killing formulation and showed that pre-training operators, derived from real fault taxonomies, tend to produce more sensitive and diverse mutants than post-training operators. 
This work established a statistically grounded methodology for measuring mutant behavior in DL systems.

More recent studies proposed operator-level quality measures to support selective mutation. 
\citet{wang_fine-grained_2023} defined per-mutant quality at the class level and aggregated it into operator-level metrics to prioritize operators. 
Their formulation remains deterministic without accounting for stochastic variation across retrainings. 
\citet{zhang_fine-grained_2025} extended this idea by aggregating per-mutant quality into operator scores and pruning low-value operators, further linking quality measurement to cost reduction. 
They also introduced fault-type diversity, but their evaluation remained operator-level, deterministic, and limited to classification tasks.

In summary, existing research on DL mutant quality either employs deterministic, single-run evaluations or probabilistic formulations that do not unify different quality dimensions. 
Our work complements these efforts by redefining mutant quality under a unified probabilistic framework that captures two complementary aspects: 
(i)~\emph{IQ}, reflecting a mutant’s resistance to killing, and 
(ii)~\emph{EQ}, reflecting its behavioral realism with respect to real faults. 
This unified view enables configuration-level selective mutation that preserves mutants that are useful for both primary roles: (i)~as a test goal, and (ii)~as fault substitution.
The following section formalizes this probabilistic framework and details how intrinsic and extrinsic qualities are computed.

\section{Methodology}
\label{sec:methodology}

\begin{figure*}[ht]
    \centering
    \includegraphics[width=0.95\linewidth]{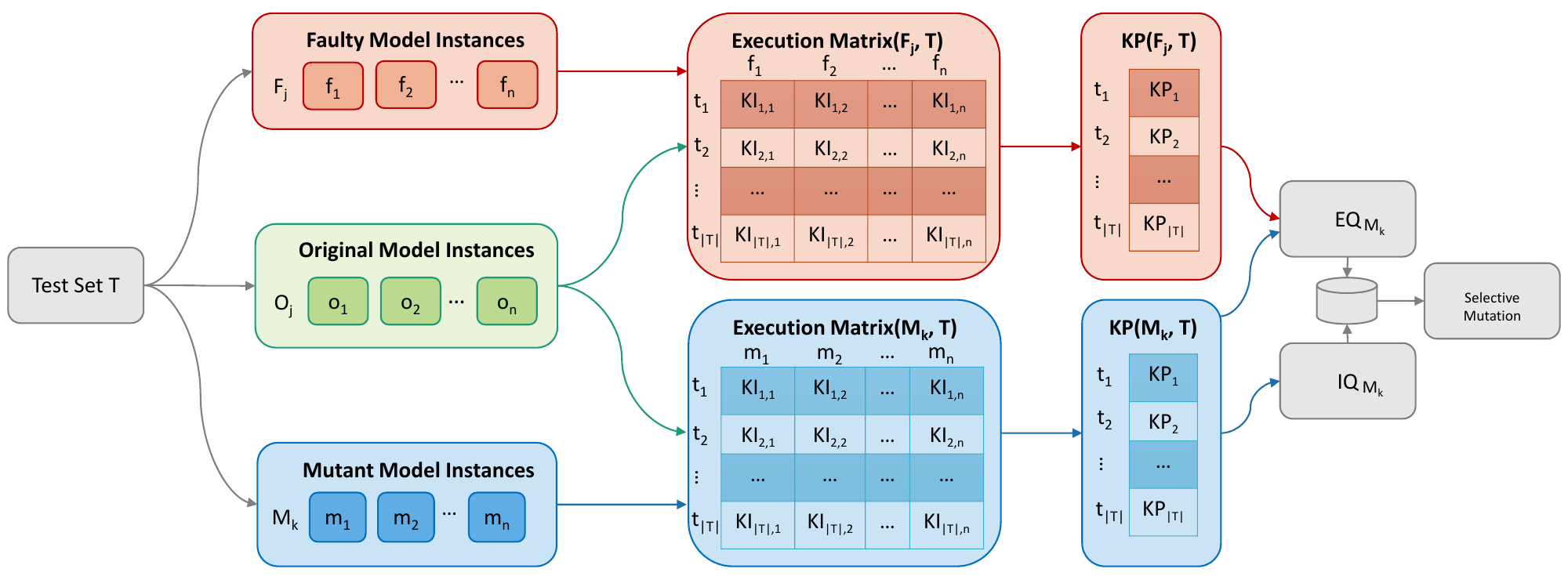}
    \caption{Overview of the methodology for assessing mutant quality using probabilistic killing in DL. KI, KP, IQ, and EQ are defined in Equations~(\ref{eq:killing-input})--(\ref{eq:eq-definition}). Adapted from \cite{ahmed_empirical_2025}.}
    \label{fig:methodology-overview}
    \Description{Overview of the workflow for computing mutant quality using probabilistic killing outcomes in deep learning models.}
\end{figure*}

This section describes the methodology used to quantify the \emph{intrinsic quality (IQ)} and \emph{extrinsic quality (EQ)} of mutants in deep learning (DL) systems. 
The process begins with constructing an \textit{execution matrix}, which captures the outcomes of running each test input against multiple independently trained instances of the same model configuration.
Because DL models exhibit stochastic behavior across independent trainings, this binary representation is extended into probabilistic form that estimates, for each test-model pair, the likelihood of behavioral deviation. 
These probabilistic data form the basis for computing two quality dimensions: 
(i)~\emph{IQ}, reflecting a mutant’s resistance to killing, and 
(ii)~\emph{EQ}, reflecting its behavioral realism relative to real faults.
The overall pipeline is illustrated in Figure~\ref{fig:methodology-overview}, adapted from our prior work \cite{ahmed_empirical_2025}.

\subsection{Model Instances}
\label{sec:model-instances}

Our methodology relies on multiple independently trained instances of \textit{original}\footnote{“Original model” follows the terminology used by the DeepCrime study for the unmutated baseline model.}, \textit{faulty}, and \textit{mutant} models to capture the stochasticity inherent in DL training. 
Following the setup established by \citet{jahangirova_empirical_2020} and extended by \citet{humbatova_deepcrime_2021}, each model type is trained multiple times under identical datasets and hyperparameter settings.
Let $n$ denote the number of retrainings per model type. 
For consistency, each faulty or mutant instance $f_i$ or $m_i$ has a corresponding original instance $o_i$ trained, enabling one-to-one comparisons between corresponding instances.

We denote the sets of trained instances as 
$O=\{o_1,\dots,o_n\}$ for the original (fault-free) models, 
$F=\{f_1,\dots,f_n\}$ for the faulty models containing real faults, and 
$M=\{m_1,\dots,m_n\}$ for the mutant models generated by pre-training mutation operators. 
All models are evaluated on a shared test dataset $T=\{t_1,\dots,t_{|T|}\}$ with ground-truth labels $y_t$. 
The resulting predictions form the raw data for the \textit{execution matrix} (Section~\ref{sec:execution_matrix}).

\subsection{Execution Matrix}
\label{sec:execution_matrix}

The \textit{execution matrix} formalizes the outcomes of executing each test input against every model instance and serves as the foundation for our quality metrics. 
The concept builds on the empirical definitions of mutation killing introduced by Jahangirova and Tonella~\cite{jahangirova_empirical_2020} and refined by Humbatova et~al.~\cite{humbatova_deepcrime_2021}, who showed that DL’s stochastic training invalidates a single deterministic notion of killing. 
Instead, killing must be interpreted statistically across multiple independent retrainings of both the original and mutated models.

Given the test dataset $T=\{t_1,\dots,t_{|T|}\}$ with ground-truth labels $y_t$, we consider paired model instances $o_i \in O$ and $x_i \in X$, where $X \in \{M, F\}$ (with $M=\{m_1,\dots,m_n\}$ denoting mutant models and $F=\{f_1,\dots,f_n\}$ denoting faulty models).
A test input $t$ is said to \textit{kill} instance $x_i$ if it is correctly classified by $o_i$ but misclassified by $x_i$, indicating a behavioral divergence between the two models.
Each entry of the \textit{execution matrix} encodes whether a test input $t$ kills a given model instance $x_i$. The corresponding \textit{killing input (KI)} value is computed as follows:

\begin{equation}
\label{eq:killing-input}
KI_{x_i}(t) =
\begin{cases}
1, & \text{if } o_i(t)=y_t \ \land\  x_i(t)\neq y_t,\\
0, & \text{otherwise.}
\end{cases}
\end{equation}

The \textit{execution matrix} is obtained by evaluating $KI_{x_i}(t)$ for all $t \in T$ and $x_i \in X$.
An entry value $KI_{x_i}(t)=1$ indicates that test input $t$ distinguishes instance $x_i$ from its corresponding original instance $o_i$ under the correctness criterion above, while $KI_{x_i}(t)=0$ indicates no distinguishing classification behavior.
Subsequent aggregation across multiple runs yields the probabilistic formulation described in Section~\ref{sec:killing_probability}, which underpins the computation of \textit{IQ} and \textit{EQ}.

\subsection{Killing Probability}
\label{sec:killing_probability}

Since training introduces randomness, the binary outcomes in the execution matrix cannot fully represent mutant detectability. 
Following the probabilistic formulation of \citet{humbatova_deepcrime_2021}, we estimate, for each test-model pair, the likelihood that a test input kills a model instance across multiple retrainings. This yields a probabilistic extension of the execution matrix, referred to as the \textit{killing probability (KP)}.

Formally, given the binary outcomes $KI_{x_i}(t)$ from $n$ independent instances belonging to $X \in \{F, M\}$ and a test input $t\in T$, the \textit{KP} is defined as:

\begin{equation}
\label{eq:killing-probability}
KP_{X}(t)=\frac{1}{n}\sum_{i=1}^{n}KI_{x_i}(t),
\end{equation}
where $x_i$ denotes the $i$-th instance of model type $X$. 
The resulting matrix $KP=[KP_{X}(t)]$ contains values in $[0,1]$, capturing the likelihood that each test input exposes behavioral deviations in faulty or mutant models.
This probabilistic representation generalizes the deterministic notion of mutant killing to account for nondeterminism and serves as the input for computing the quality measures presented next.

\subsection{Intrinsic Quality}
\label{sec:intrinsic_quality}

The \textit{IQ} quantifies a mutant’s resistance to being killed, reflecting how difficult it is for test inputs to expose its behavioral deviations.
This concept originates from the classical notion of \textit{resistant mutants}~\citep{estero-botaro_quality_2015}, where the quality of a mutant depends on how many tests kill it and on how many other mutants are killed by those same tests.
Prior work has adapted this notion to DL under deterministic killing assumptions by constructing adequate and non-redundant test suites~\citep{zhang_fine-grained_2025}.
In contrast, we adapt the underlying idea of mutant quality to the probabilistic and statistical nature of DL systems, without assuming adequacy or non-redundancy of the test set.

For \textit{IQ}, we restrict attention to mutant models and write $KP_m(t)$ as a shorthand for $KP_X(t)$ when $X = M$.
Let $M$ denote the set of generated mutants, $T$ the test set, and $KP_m(t)$ the probability that test input $t$ kills mutant $m$, as defined in Section~\ref{sec:killing_probability}.
For each test input $t \in T$, we define its probabilistic mutant coverage as:

\begin{equation}
C_t = \frac{1}{|M|}\sum_{m' \in M} KP_{m'}(t),
\end{equation}
where $C_t \in [0,1]$ captures how generic or discriminating a test input is. A value close to~1 indicates that the input strongly kills many mutants, whereas a value close to~0 indicates weak or rare killing behavior.

For each mutant $m$, we further define its average killing probability across the test set as:

\begin{equation}
S_m = \frac{1}{|T|} \sum_{t \in T} KP_m(t),
\end{equation}

\noindent
which quantifies how easily the mutant is killed overall.
The \textit{IQ} of a mutant $m$ is then defined as:
\begin{equation}
\label{eq:iq-definition}
IQ_m =
\begin{cases}
0, & \text{if } S_m = 0, \\
\bigl(1 - S_m\bigr)
\left(
1 - \dfrac{\sum_{t \in T} KP_m(t)\, C_t}
{\sum_{t \in T} KP_m(t)}
\right),
& \text{otherwise.}
\end{cases}
\end{equation}

This definition captures two complementary aspects of mutant quality.
The first factor, $1 - S_m$, reflects the mutant’s resistance to killing, assigning higher quality to mutants that are not easily killed across the test set.
The second factor measures whether the inputs that kill the mutant are discriminating, penalizing mutants that are primarily killed by inputs that also kill many other mutants.
This formulation generalizes the deterministic quality metric of \citet{estero-botaro_quality_2015} to a statistical context, where each test contributes proportionally to the probability of killing rather than as a binary outcome.

\subsection{Extrinsic Quality}
\label{sec:extrinsic_quality}

The \textit{EQ} measures a mutant’s realism by comparing its detectability pattern with that of the corresponding real fault. 
Let $KP_m(t)$ and $KP_f(t)$ denote the killing probabilities (Section~\ref{sec:killing_probability}) for mutant $m\in M$ and its paired faulty model $f\in F$ on each test input $t\in T$. 
Each mutant is compared to the faulty model of the same subject system, using the corresponding original model as a common behavioral baseline.
Following the generalized Jaccard similarity for fuzzy sets~\citep{castro_fernandez_lazo_2019}, we compute the similarity between their detectability profiles as:

\begin{equation}
\label{eq:eq-definition}
EQ_m =
\frac{\sum_{t\in T}\min\{KP_m(t),\,KP_f(t)\}}
     {\sum_{t\in T}\max\{KP_m(t),\,KP_f(t)\}},
\end{equation}

\noindent
where values range from 0 (no overlap) to 1 (identical detectability). 
Higher $EQ_m$ indicates that the mutant exhibits detectability patterns similar to its corresponding real fault, implying greater behavioral realism. 
Lower values suggest divergent behavior, making the mutant less representative of the real fault.

\section{Experiment Design}
\label{sec:experiment_design}
This section describes the experimental design used to evaluate the proposed quality measures.

\subsection{Research Questions}
\label{sec:research_questions}

This study investigates how IQ and EQ measures can be used to analyze and optimize mutant generation in DL systems.
Based on the motivations outlined in Section~\ref{sec:introduction} and the methodological framework in Section~\ref{sec:methodology}, we formulate the following research questions:

\begin{itemize}
    \item \textbf{RQ1:} How does IQ characterize the resistance of DL mutants to being killed across mutation operators?
    
    \item \textbf{RQ2:} How does EQ characterize the behavioral realism of DL mutants across mutation operators?
    
    \item \textbf{RQ3:} What is the relationship between IQ and EQ of DL mutants?
    
    \item \textbf{RQ4:} Can IQ and EQ be used to identify a reduced set of mutation operator configurations while preserving both resistance and behavioral realism?
\end{itemize}

\noindent
RQ1 focuses on assessing IQ as a probabilistic measure of mutant resistance, characterizing how difficult mutants are to be killed across different mutation operators and their configurations.
RQ2 analogously focuses on EQ, analyzing how mutation operators influence the behavioral realism of mutants with respect to real faults.
Together, RQ1 and RQ2 provide complementary operator-wise views of resistance and realism.

While RQ1 and RQ2 analyze resistance and realism independently, RQ3 investigates the relationship between them at the mutant level, examining whether highly resistant mutants also tend to exhibit realistic detectability behavior or whether systematic trade-offs emerge between IQ and EQ.
RQ4 builds on the insights from RQ3 to evaluate whether IQ and EQ can be used to guide a quality-driven selection of mutation operator configurations.

\subsection{Datasets}
\label{sec:datasets}
We use four datasets of real DL faults with paired faulty and fixed (original) model versions.
Mutants are generated from original models using the pre-training mutation operators listed in Table~\ref{tab:operators}, and models are retrained multiple times ($n = 5$ in our experiments) to enable probabilistic killing analysis.

Three datasets, CleanML~\citep{li_cleanml_2021}, DeepFD~\citep{cao_deepfd_2022}, and DeepLocalize~\citep{wardat_deeplocalize_2021}, serve as selection datasets for RQ1--RQ3 and cover diverse fault types and model behaviors.
Defect4ML~\citep{morovati_bugs_2023} is held out exclusively for RQ4 validation to assess whether quality-driven configuration selection reduces mutant count while preserving median IQ and EQ on unseen faults.
Bug selection follows reproducibility criteria from prior work~\citep{ahmed_empirical_2025}.
Table~\ref{tab:datasets-analysis} summarizes the included faults.

\begin{table}[h]
  \centering
  \caption{Summary of Datasets and Faults}
  \label{tab:datasets-analysis}
  \begin{tabular}{lcc}
    \toprule
    Dataset & Total Faults & Included Faults \\
    \midrule
    CleanML~\citep{li_cleanml_2021} & 23 & 19 \\
    DeepFD~\citep{cao_deepfd_2022} & 58 & 25 \\
    DeepLocalize~\citep{wardat_deeplocalize_2021} & 40 & 18 \\
    defect4ML~\citep{morovati_bugs_2023} & 100 & 24 \\
    \bottomrule
  \end{tabular}
\end{table}

\subsection{Mutation Operators}
\label{sec:mutation_operators}

We use the pre-training mutation operators defined in DeepCrime \citep{humbatova_deepcrime_2021}, which introduce changes to training data, model architecture, and training hyperparameters before model training.
Each operator introduces a specific type of change and is parameterized to generate multiple configurations, corresponding to different parameter settings and target components.
All operators, parameters, and configuration space used in this study are listed in Table~\ref{tab:operators}.

\begin{table*}
\centering
\scriptsize
\caption{Pre-training mutation operators with parameters and configuration spaces}
\label{tab:operators}
\begin{tabular}{p{0.04\linewidth} p{0.17\linewidth} p{0.23\linewidth} p{0.45\linewidth}}
\toprule
\textbf{ID} & \textbf{Operator} & \textbf{Parameters} & \textbf{Configuration Space} \\
\midrule
TCL & Change labels of training data & Label, percentage of data & Label: most frequent in training data as default or custom; Percentage: $(0, 100]$, binary search\\
TRD & Remove portion of training data & Percentage of data & Percentage: $(0, 99]$, binary search \\
TUD & Unbalance training data & Percentage of data & Percentage: $(0, 100]$, binary search \\
TAN & Add noise to training data & Percentage of data & Percentage: $(0, 100]$, binary search \\
TCO & Make output classes overlap & Percentage of data & Percentage: $(0, 100]$, binary search \\
HBS & Change batch size & New batch size & Batch size: \{$b/4$, $b/2$, $2b$, $4b$\} where $b$ is original batch size, exhaustive \\
HLR & Decrease learning rate & New learning rate & Learning rate: ($\sim$0, $lr$], binary search with precision $lr/10$ where $lr$ is original value \\
HNE & Change number of epochs & New number of epochs & Epochs: [1, $epochs$], binary search with precision $epochs/10$ \\
HDB & Disable data batching & --- & Binary toggle: only if batching enabled \\
ACH & Change activation function & Layer number, new activation function & Layer: all with non-linear activation; Function: all Keras activations except linear, exhaustive \\
ARM & Remove activation function & Layer number & Layer: all with non-linear activation, exhaustive \\
AAL & Add activation function to layer & Layer number, new activation function & Layer: all with linear activation; Function: all Keras activations except linear, exhaustive \\
RAW & Add weights regularisation & Layer number, new weights regularisation & Layer: all without regularisation; Regularisation: \{L1, L2, L1L2\}, exhaustive \\
RCW & Change weights regularisation & Layer number, new weights regularisation & Layer: all with existing regularisation; Regularisation: \{L1, L2, L1L2\}, exhaustive \\
RRW & Remove weights regularisation & Layer number & Layer: all with existing regularisation, exhaustive \\
RCD & Change dropout rate & Layer number, new dropout rate & Layer: all dropout layers; Rate: 4 values in $(0, 1]$ excluding original rate, exhaustive \\
RCP & Change patience parameter & New patience value & Patience value: [$1, patience$], binary search with precision $patience/10$ \\
WCI & Change weights initialisation & Layer number, new initialisation & Layer: all applicable layers; Initialiser: all Keras initialisers, exhaustive \\
WAB & Add bias to a layer & Layer number & Layer: all layers without bias, exhaustive \\
WRB & Remove bias from a layer & Layer number & Layer: all layers with bias, exhaustive \\
LCH & Change loss function & New loss function & Loss function: all Keras loss functions except original, exhaustive \\
OCH & Change optimisation function  & New optimisation function & Optimisation function: all Keras optimisers except original, exhaustive \\
OCG & Change gradient clipping & New gradient clipping & Gradient clipping: custom values, exhaustive \\
VRM & Remove validation set & --- & Binary toggle: only if validation set is used \\
\bottomrule
\end{tabular}
\end{table*}

Each configuration fixes a mutation operator and its parameter values, forming a unique pre-training mutation setting.
Applying a mutation operator with a specific configuration to a model produces one mutant.

\subsection{Evaluation Procedure}
\label{sec:evaluation_procedure}
The evaluation follows the probabilistic workflow defined in Section~\ref{sec:methodology}.
For each bug, all mutation operator configurations (Table~\ref{tab:operators}) are applied to the original model, yielding one mutant per configuration.
Each original, faulty, and mutant model is retrained five times under identical settings.
Prediction outcomes are collected to construct execution matrices, from which killing probabilities and subsequently IQ and EQ values are computed (Sections~\ref{sec:execution_matrix}--\ref{sec:extrinsic_quality}).

For the selection datasets (CleanML, DeepFD, DeepLocalize), mutant-level IQ and EQ values support RQ1-RQ3 through operator-wise and joint analyses.
For RQ4, raw configurations are mapped to canonical families that abstract model-specific parameters (e.g., \texttt{ACH\_relu\_layer\_3} → \texttt{ACH\_relu}).
Each family's hit rate is computed as the proportion of mutants whose IQ and EQ both exceed dataset-specific medians.
Families exceeding threshold $\tau$ are retained.
No information from the held-out dataset (defect4ML) is used during selection.
Retained families are applied to defect4ML to evaluate effectiveness by comparing mutant counts and IQ/EQ distributions before and after selection.
All experiments are automated under identical environments.

\subsection{Evaluation Metrics}
\label{sec:evaluation_metrics}

We evaluate the approach using quantitative metrics corresponding to each research question.

\paragraph{RQ1--RQ2: Characterizing IQ and EQ}
For each mutation operator, we analyze distributions of mutant-level IQ and EQ values across the selection datasets using medians and interquartile ranges.

\paragraph{RQ3: Relationship Between IQ and EQ}
Mutants are examined in the two-dimensional IQ--EQ space using dataset-specific median-based thresholds, partitioning the space into four quality quadrants for systematic comparison.

\paragraph{RQ4: Configuration Selection and Cost Reduction}
Cost reduction is quantified using the reduction ratio:
\[
RR = 1 - \frac{|\text{mutants after selection}|}{|\text{mutants before selection}|}.
\]
This metric captures the relative decrease in the number of generated mutants after applying configuration selection.
Quality impact is assessed using two indicators: (i) changes in the median IQ and EQ values, and (ii) changes in the proportion of mutants whose IQ and EQ simultaneously exceed the dataset-specific medians, measured before and after selection on the held-out dataset.

\section{Results and Analysis}
\label{sec:results_analysis}
This section presents the empirical results of our study and analyzes them with respect to the defined research questions.

\subsection{RQ1: IQ of DL Mutants}
\label{sec:results_rq1}

\begin{figure*}[ht]
    \centering
    \includegraphics[width=\linewidth]{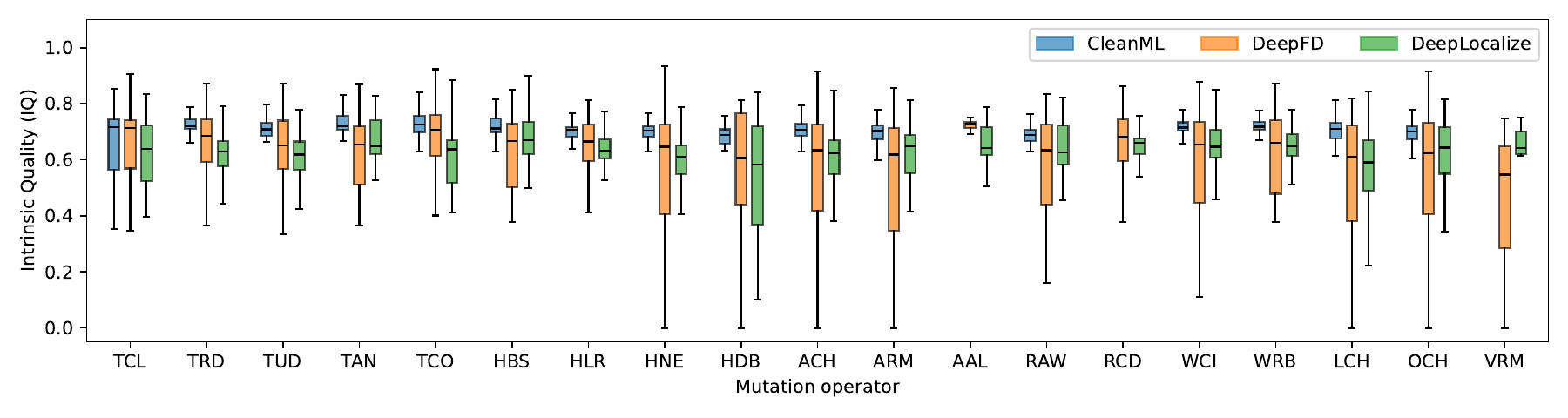}
    \caption{Operator-wise distributions of mutant-level IQ across CleanML, DeepFD, and DeepLocalize, grouped by dataset.}
    \label{fig:operator_iq_boxplot_by_dataset}
    \Description{Box-plots illustrating the varying IQ of operator-wise mutants}
\end{figure*}

Figure~\ref{fig:operator_iq_boxplot_by_dataset} shows the distribution of \(IQ\) values for mutants generated by each pre-training mutation operator.
Each boxplot aggregates mutant-level \(IQ\) values across all configurations of the operator and across all subject systems in the selection datasets (CleanML, DeepFD, and DeepLocalize).

Across all operators, \(IQ\) exhibits substantial dispersion rather than concentrating around a narrow range.
For most operators, the interquartile ranges span values roughly between 0.50 and 0.75, while lower whiskers often extend toward lower values, particularly for mutants generated from DeepFD.
The wider IQ spread observed for DeepFD indicates greater variability in how different mutation operators affect mutant resistance on this dataset, compared to the others.
In contrast, the more compact IQ distributions observed for CleanML can be attributed to the use of the same multilayer perceptron classifier across faults within the dataset.
This constrains variation in prediction behavior and, in turn, limits the variability captured by the prediction-based IQ measure.
Notably, no operator yields mutants with uniformly high or uniformly low \(IQ\) values across all configurations and datasets.

These results demonstrate that \(IQ\) captures meaningful variation among mutants generated by the same mutation operator across configurations and subject systems.
This supports the use of \(IQ\) as a probabilistic measure for characterizing mutant resistance in DL systems.

\subsection{RQ2: EQ of DL Mutants}
\label{sec:results_rq2}

\begin{figure*}[ht]
    \centering
    \includegraphics[width=1.0\linewidth]{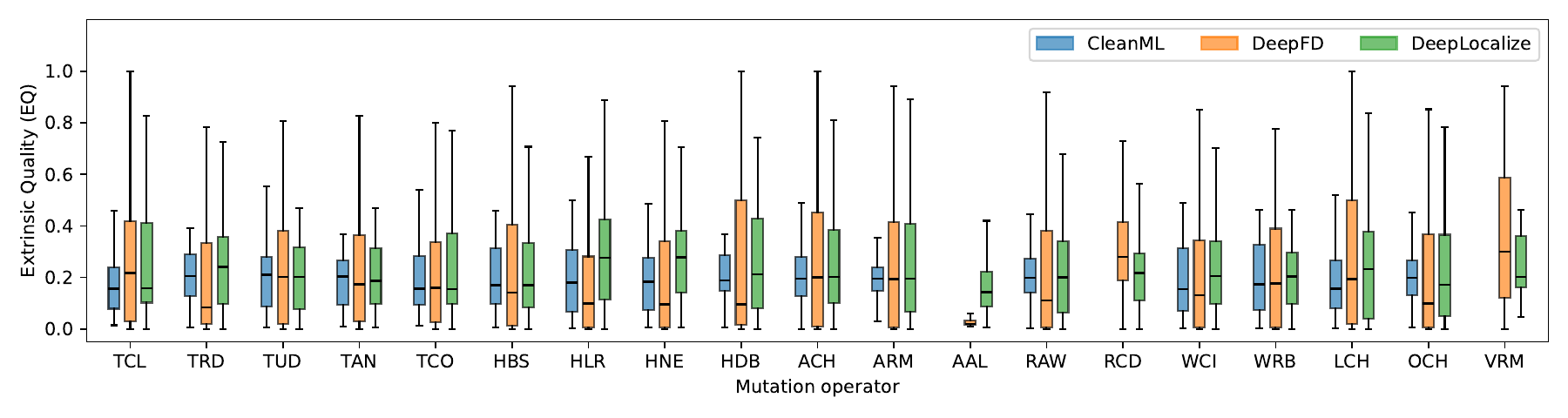}
    \caption{Operator-wise distributions of mutant-level EQ across CleanML, DeepFD, and DeepLocalize, grouped by dataset.}
    \label{fig:operator_eq_boxplot_by_dataset}
    \Description{Box-plots illustrating the varying EQ of operator-wise mutants}
\end{figure*}

Figure~\ref{fig:operator_eq_boxplot_by_dataset} shows the distribution of \(EQ\) values for mutants generated by each pre-training mutation operator.
Each boxplot aggregates mutant-level \(EQ\) values across all configurations of the operator and across all bug--fix pairs in the selection datasets (CleanML, DeepFD, and DeepLocalize).

Across operators and datasets, median \(EQ\) values are generally low, typically ranging between approximately 0.15 and 0.30.
This indicates that most mutants exhibit only partial overlap with the detectability behavior of the corresponding real faults.
At the same time, the distributions consistently show long upper tails, with many operators producing mutants whose \(EQ\) values exceed 0.6 and, in some cases, approach 1.0.
These high-\(EQ\) values indicate that certain configurations are capable of generating mutants whose detectability behavior closely resembles that of real faults.
The dispersion of \(EQ\) values varies across datasets.
Mutants associated with DeepFD exhibit broader \(EQ\) distributions with more pronounced upper ranges, whereas CleanML and DeepLocalize show comparatively more compact distributions.
Nevertheless, for all datasets, no mutation operator yields mutants with uniformly high \(EQ\) values across all configurations.

Overall, these results show that while most mutants exhibit limited behavioral realism, \(EQ\) reveals a clear separation between weakly and highly realistic mutants through its pronounced upper tails.
This supports the use of \(EQ\) for identifying mutation configurations that generate mutants closely resembling the detectability behavior of real faults.

\subsection{RQ3: Relationship Between IQ and EQ}
\label{sec:results_rq3}

\begin{figure*}[ht]
    \centering
    \includegraphics[width=1.0\linewidth]{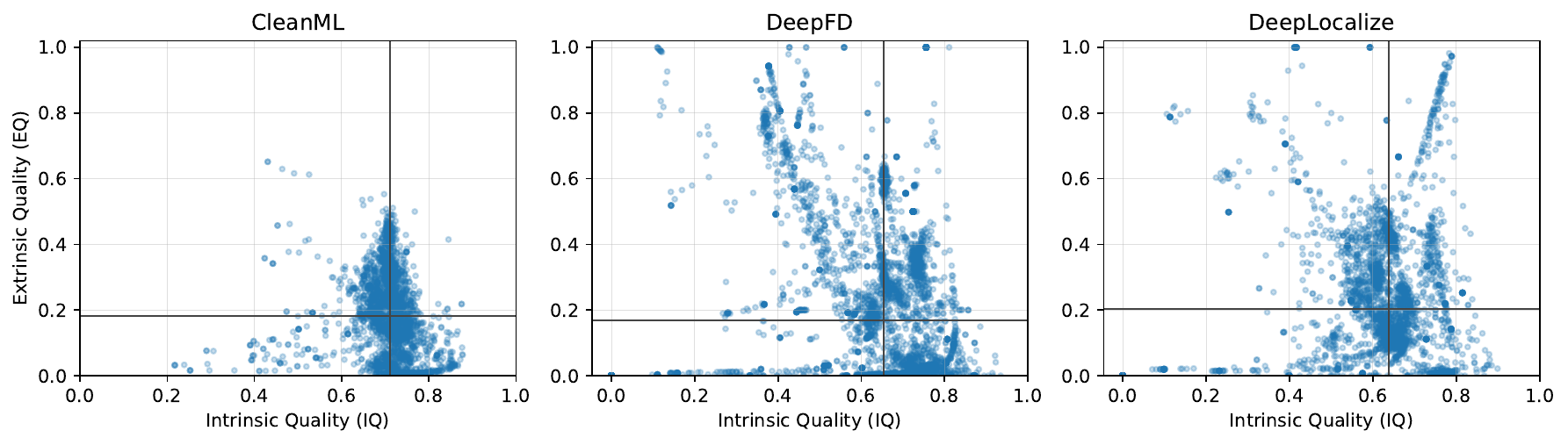}
    \caption{Mutant-level IQ--EQ scatter with median-based quadrant partitioning, illustrating the relationship between resistance and realism.}
    \label{fig:quadrant_scatter_by_dataset}
    \Description{Scatter plot of mutant-level intrinsic quality (IQ) versus extrinsic quality (EQ), partitioned into four quadrants using median-based thresholds to illustrate the relationship between resistance to killing and behavioral realism.}
\end{figure*}

Figure~\ref{fig:quadrant_scatter_by_dataset} illustrates the joint distribution of \(IQ\) and \(EQ\) for all mutants in the selection datasets.
The four quadrants correspond to combinations of high and low resistance and realism.
Although the quadrants are defined using median thresholds, the distribution of mutants across quadrants varies across datasets, suggesting that high or low values of \(IQ\) do not systematically coincide with high or low values of \(EQ\).

Across all datasets, mutants are distributed across all four quadrants, with the relative proportions varying by dataset.
For CleanML, approximately 19\% of mutants fall into the high-\(IQ\), high-\(EQ\) quadrant, while the remaining mutants are spread across the mixed and low-quality regions.
DeepFD exhibits an almost uniform distribution across quadrants, with roughly one quarter of mutants in each region.
A similar pattern is observed for DeepLocalize, where the proportions of mutants in each quadrant differ only marginally.
These distributions suggest that intrinsic resistance and extrinsic realism capture different aspects of mutant quality.
High \(IQ\) does not imply high \(EQ\), and realistic mutants are not necessarily difficult to kill.
At the same time, a non-trivial proportion of mutants consistently achieve high values for both metrics across all datasets.

Overall, the quadrant analysis demonstrates that \(IQ\) and \(EQ\) capture complementary dimensions of mutant quality.
The presence of a substantial minority of mutants in the high-\(IQ\), high-\(EQ\) quadrant indicates that it is possible to generate mutants that are both resistant and behaviorally representative of real faults.
This observation provides empirical support for using joint quality criteria when selecting mutation operator configurations, which is explored in RQ4.

\subsection{RQ4: Operators Configuration Selection and Validation}
\label{sec:results_rq4}

RQ4 investigates whether IQ and EQ can guide a selective mutation strategy that reduces the number of generated mutants while preserving resistance and behavioral realism.
We address RQ4 in two steps: (i) configuration selection using the selection datasets (CleanML, DeepFD, DeepLocalize) and (ii) validation on the held-out dataset (defect4ML).

\subsubsection{Selection on CleanML, DeepFD, and DeepLocalize}
A direct union of raw configuration strings is not stable across subject systems because several mutation operators encode model-dependent parameters (e.g., layer indices for layer-level operators or discrete percentage values for data mutation operators).
To ensure transferability and reproducibility, we map raw configurations to \emph{canonical configuration families}, which abstract away model-specific details while retaining the semantic intent of the mutation operator and coarse-grained parameter choices.
For instance, layer-level operators such as ARM produce configurations like \texttt{ARM\_layer\_3} and \texttt{ARM\_layer\_7}, which are mapped to the same canonical family \texttt{ARM}, removing layer-specific indices.
Similarly, percentage-based operators such as TRD generate configurations \texttt{TRD\_pct\_8}, \texttt{TRD\_pct\_12}, and \texttt{TRD\_pct\_15}, which are canonicalized to \texttt{TRD\_pct\_5\_15}, grouping configurations with similar parameter magnitudes into a shared family.

Selection statistics are therefore computed per canonical configuration family rather than per raw configuration string.
Using this abstraction, configuration families are evaluated across the selection datasets (CleanML, DeepFD, and DeepLocalize).
We use a hit-rate–based criterion to quantify how frequently a configuration family produces high-quality mutants across datasets.
Table~\ref{tab:selected_operators} summarizes the mutation operators for which at least one configuration family is retained, together with their selected canonical configuration spaces.
Ten of the 24 mutation operators listed in Table~\ref{tab:operators} are excluded from further analysis: five (HNE, HDB, ARM, AAL, RAW) are filtered during quality-driven selection under hit-rate threshold $\tau=0.25$, while five (RCW, RRW, RCP, WAB, OCG) are not applicable to any subject system.

Our selection criterion follows the median-based quadrant partitioning introduced in RQ3 (Figures~\ref{fig:quadrant_scatter_by_dataset}).
For each selection dataset, we compute dataset-specific medians of IQ and EQ.
A mutant is labeled \emph{High--High} if both its IQ and EQ are at least the corresponding medians, placing it in the top-right quadrant of the IQ--EQ space.
For each canonical configuration family, we compute its \emph{hit-rate} as the proportion of generated mutants that fall into the High--High quadrant across the selection datasets.
A configuration family is retained if its hit rate exceeds a threshold $\tau$, as justified below.

\begin{table}
\centering
\scriptsize
\caption{Retained mutation operators with selected configuration spaces under hit-rate threshold $\boldsymbol{\tau=0.25}$}
\label{tab:selected_operators}
\begin{tabular}{p{0.07\linewidth} p{0.85\linewidth}}
\toprule
\textbf{ID} & \textbf{Selected Configuration Space} \\
\midrule
TCL & Label: default or custom; Percentage: $(0,30] \cup (50,90]$, binary search \\
TRD & Percentage: $(5,90]$, binary search \\
TUD & Percentage: $(0,5] \cup (30,70]$, binary search \\
TAN & Percentage: $(5,50] \cup (70,100]$, binary search \\
TCO & Percentage: $(0,30] \cup (50,90]$, binary search \\
HBS & Batch size: $b' \in (32,256] \cap \{b/4,b/2,2b,4b\}$, exhaustive\\
HLR & Learning rate: ($0.01$, $lr$], binary search with precision $lr/10$ \\
ACH & Layer: all with non-linear activation; Function: $A' \in \{\texttt{relu},\,\texttt{selu}\}\setminus\{A\}$, exhaustive \\
RCD & Layer: all dropout layers; Rate: $r' \in \{0.125,\,0.25,\,0.75\} \setminus \{r\}$, exhaustive \\
WCI & Layer: all applicable; Initializer: all Keras $\setminus\{\texttt{zeros},\,\texttt{ones},\,\texttt{constant}\}$, exhaustive \\
WRB & Layer: all layers with bias, exhaustive \\
LCH & Loss function: $L' \in \{\texttt{CCE},\,\texttt{MSLE},\,\texttt{COS}\}\setminus\{L\}$, exhaustive \\
OCH & Optimisation function: $O' \in \{\texttt{nadam},\,\texttt{rmsprop}\}\setminus\{O\}$, exhaustive \\
VRM & Binary toggle: only if validation set is used \\
\bottomrule
\end{tabular}
\end{table}

\paragraph{Rationale for the Hit-Rate Selection Criterion ($\tau = 0.25$).}
The hit-rate threshold $\tau$ specifies the minimum frequency with which a canonical configuration family produces \emph{High--High} mutants.
High--High means that both \(IQ\) and \(EQ\) are at least the dataset-specific medians under the median-based quadrant partitioning of RQ3.
Across the selection datasets, the High--High proportion is typically around one quarter, ranging approximately between \(0.20\) and \(0.25\).
We use this range as an empirical reference for a non-trivial concentration of jointly high-quality mutants.
We set $\tau = 0.25$ as a simple cutoff at the upper end of this range.
A configuration family is retained only if at least one quarter of its mutants are High--High.
This avoids retaining families that reach High--High only occasionally.

To assess robustness, we additionally evaluated nearby thresholds ($\tau = 0.20$ and $\tau = 0.30$).
As $\tau$ increases, the number of retained configuration families decreases monotonically, reflecting a clear and interpretable cost--quality trade-off.
Specifically, $\tau = 0.20$ retains $67$ out of $123$ families ($\mathrm{RR} \approx 46\%$), $\tau = 0.25$ retains $51$ families ($\mathrm{RR} \approx 59\%$), and $\tau = 0.30$ retains $27$ families ($\mathrm{RR} \approx 78\%$).
Lower thresholds retain more families but require a lower High--High frequency, while higher thresholds yield stronger reduction at the cost of excluding more families.
Across these settings, the selection outcome changes predictably with $\tau$, showing a monotonic trade-off between configuration reduction and selection strictness.

\subsubsection{Validation on defect4ML}
\begin{figure}[ht]
    \centering
    \includegraphics[width=1.0\linewidth]{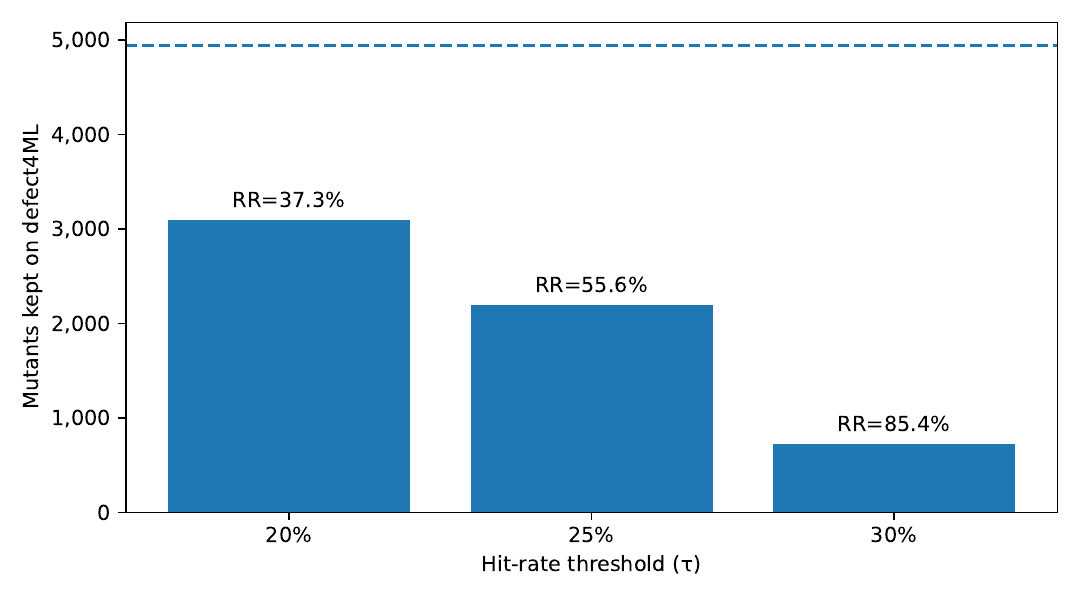}
    \caption{Number of mutants retained on defect4ML under different hit-rate thresholds.}
    \label{fig:mutants_kept_by_threshold}
    \Description{Bar chart showing the number of mutants retained on defect4ML at different hit-rate thresholds, with a dashed line indicating the baseline.}
\end{figure}

We validate the selected configuration families on defect4ML, which is not used during the selection process.
This validation assesses whether a selection rule derived from CleanML, DeepFD, and DeepLocalize generalizes to an unseen dataset.
The baseline defect4ML experiment includes $4{,}939$ pre-training mutants generated using all applicable mutation operator configurations.
After applying configuration-family selection, we regenerate only those mutants whose canonical families are retained and compare mutant generation cost (measured by mutant count) and quality before and after selection.
Figure~\ref{fig:mutants_kept_by_threshold} reports the number of mutants retained under different hit-rate thresholds.
Selection yields substantial and monotonic reductions in mutant count.
In particular, with $\tau = 0.25$, the number of mutants decreases from $4{,}939$ to $2{,}194$, corresponding to a reduction ratio of $55.6\%$.
Higher thresholds lead to stronger reductions, consistent with the behavior observed during selection.

\begin{figure}[ht]
    \centering
    \includegraphics[width=1.0\linewidth]{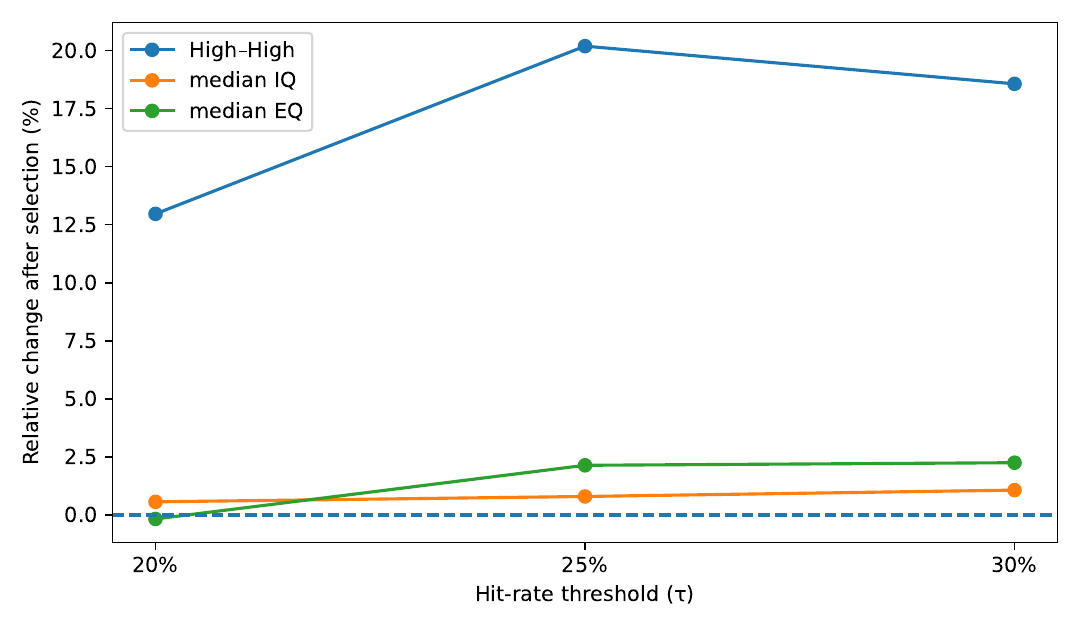}
    \caption{Relative changes in  High--High proportion, median IQ, and median EQ on defect4ML after selection.}
    \label{fig:relative_change_percent}
    \Description{Line plot showing relative changes in median IQ, median EQ, and the High--High mutant proportion on defect4ML across hit-rate thresholds, with a dashed zero-change reference line.}
\end{figure}

To assess the impact of selection on mutant quality, we examine both the preservation of central tendency and the enrichment of jointly high-quality mutants.
Concretely, we compare the median \(IQ\) and \(EQ\) values before and after selection and measure the change in the proportion of High--High mutants.
In the defect4ML baseline, the proportion of High--High mutants is $0.221$.
After selection with $\tau = 0.25$, this proportion increases to $0.266$, indicating that the retained configuration families generate a higher concentration of mutants that are jointly high in resistance and realism.
At the same time, the medians of IQ and EQ change only marginally after selection, indicating that typical mutant quality is preserved despite the reduction in mutant count.
Figure~\ref{fig:relative_change_percent} summarizes these effects as relative changes, showing that selection primarily alters quadrant proportions rather than shifting central tendency.

This behavior is expected because selection operates at the level of configuration families and is driven by frequency rather than magnitude: families that rarely produce jointly high-quality mutants are removed, while the overall scale of IQ and EQ values is preserved.

Overall, these results show that IQ and EQ can be used to define a practical and transferable selective mutation strategy.
Configuration-family selection substantially reduces the cost of mutant generation on unseen faults while increasing the proportion of mutants that are jointly resistant and behaviorally realistic under the same median-based quality definition used throughout the study.

\section{Discussion}
\label{sec:discussion}

This study set out to rethink selective mutation for DL from a quality-oriented perspective rather than an operator-centric one. While Section~\ref{sec:results_analysis} established the empirical behavior of the proposed quality measures, this section reflects on their broader meaning, limitations, and implications.

\subsection{Resistance and Realism as Independent Quality Dimensions}

A central conceptual insight emerging from RQ1–RQ3 is that resistance to killing and behavioral realism represent independent dimensions of mutant quality.
RQ1 shows that IQ varies substantially across configurations, even within the same mutation operator, indicating that resistance is sensitive to parameter choices rather than being an operator-level constant.
RQ2 complements this observation by showing that EQ varies independently across operators, reflecting how different mutation mechanisms align, or fail to align, with the detectability patterns of real faults. 

The quadrant-based analysis in RQ3 makes this separation explicit: mutants populate all regions of the IQ--EQ space, and no simple monotonic relationship emerges between resistance and realism. Taken together, these findings confirm that optimizing mutants solely for resistance, or solely for realism, would overlook important parts of the mutant space and lead to an incomplete characterization of mutant quality.

\subsection{Configuration Selection and Generalization}

The configuration-family selection strategy shows how the proposed quality measures support selective mutation across different subject systems.
A key challenge in DL mutation testing is that raw configuration strings often encode model-specific parameters, such as layer indices or baseline-dependent hyperparameter values.
These details prevent direct transfer across models.
By mapping raw configurations to canonical configuration families that preserve the semantic choice of an operator while abstracting away model-dependent details, selection rules derived from representative datasets can be applied to unseen systems.

The hit-rate criterion operationalizes quality-driven selection at the configuration-family level.
Selection is based on how consistently a family produces mutants that are jointly high in resistance and realism, rather than on isolated high-quality outliers. This frequency-based filtering favors configuration families with stable quality patterns across datasets.
The validation on defect4ML shows that selection derived using CleanML, DeepFD, and DeepLocalize transfers effectively to unseen faults.
It achieves substantial cost reduction while enriching High–High mutant proportion without materially shifting median IQ or EQ.

Dataset-specific medians further support transferability.
Defining high and low regions relative to observed quality distributions normalizes IQ and EQ across datasets with different scales.
This avoids calibration issues associated with fixed global thresholds.
As a result, selection mainly affects quadrant proportions rather than shifting central tendency, which aligns with the goal of removing low-value configurations while preserving representative mutant behavior.

\subsection{Limits of Quality-Driven Selection}

Despite its benefits, quality-driven selection does not address all challenges inherent in selective mutation for DL.
IQ and EQ focus on resistance and detectability behavior, and therefore characterize how mutants behave under testing rather than why they behave that way. 
In particular, EQ captures behavioral similarity in terms of overlapping killing patterns but does not establish causal or structural correspondence.
Mutants with high EQ may resemble real faults in detectability while differing in underlying fault mechanisms. 
As a result, quality-driven selection should be viewed as complementary to, rather than a replacement for, approaches that explicitly target fault causation or diversity of mutation effects.

Moreover, prioritizing high-quality regions of the IQ--EQ space may reduce exposure to low-frequency but potentially informative mutants.
While such mutants may be less attractive from a quality perspective, they can still reveal edge-case weaknesses in specific testing scenarios.
Balancing quality-driven reduction with sufficient behavioral coverage remains an open problem.

\subsection{Implications}

The findings of this study have several implications for both research and practice.

\subsubsection{Implications for Research}
The results show that separating resistance from realism benefits selective mutation for DL systems.
By quantifying these properties independently, IQ and EQ provide general-purpose quality measures that extend beyond configuration selection.
They can be used to rank mutants, operators, or configurations according to different testing objectives and evaluation budgets.
Rather than treating mutation operators as uniform generators, this study shows that operators induce quality distributions across configurations.
This perspective enables systematic comparison and supports the design of quality-aware mutation strategies.
More broadly, IQ and EQ offer a foundation for studying cost--quality trade-offs, where low-value mutants can be deprioritized before expensive executions.

\subsubsection{Implications for Practice}
For practitioners conducting large-scale DL mutation experiments, the proposed approach offers a practical way to control cost without sacrificing testing relevance.
Mutant generation can be restricted to configuration families that empirically yield higher concentrations of jointly resistant and realistic mutants.
The hit-rate threshold provides an explicit control parameter to balance computational budget and quality under different evaluation constraints.

\section{Threats to Validity}
\label{sec:threats}

\paragraph{Internal validity.}
DL training is stochastic due to random initialization, data shuffling, and non-deterministic GPU operations.
To mitigate this, we retrain each model five times under identical settings and compute killing probabilities from repeated runs rather than single executions.
Some variance remains unavoidable and can slightly affect estimated IQ, EQ, and quadrant assignments.

\paragraph{Construct validity.}
IQ and EQ are operational measures for resistance and realism.
IQ depends on killing probabilities and test weighting via coverage term $C_t$.
Near-duplicate inputs may inflate common killing patterns, a realistic limitation we acknowledge.
Our quadrant analysis uses dataset-specific medians to define High and Low regions, avoiding fixed thresholds that may not transfer across datasets with different IQ/EQ scales.
The hit-rate threshold $\tau=0.25$ is justified by observed High--High baselines (Figure~\ref{fig:quadrant_scatter_by_dataset}) and validated across multiple thresholds.

\paragraph{External validity.}
Our experiments focus on classification tasks using Keras/TensorFlow due to tool compatibility and dataset availability.
Selected configuration families may not transfer to regression tasks, other frameworks (e.g., PyTorch), or post-training mutation operators.
Our sample comprises 86 reproducible bug-fix pairs from four datasets, with 62 for selection and 24 for held-out validation.
This reflects practical constraints on reproducible DL faults and supports clear train-test separation.
Results generalize to DL faults that can be reliably executed in controlled settings.

\paragraph{Conclusion validity.}
We use descriptive analysis and cross-dataset validation rather than extensive significance testing.
Future work could add confidence intervals or hypothesis tests to strengthen inferential claims.

\section{Conclusion and Outlook}
\label{sec:conclusion}

This paper proposed a quality-driven approach to selective mutation for DL based on two complementary measures: IQ for resistance and EQ for behavioral realism.
Using a probabilistic killing framework, we showed that IQ characterizes mutant resistance across operator configurations (RQ1), EQ characterizes behavioral similarity to real faults (RQ2), and these dimensions are largely independent (RQ3).
Building on these findings, we introduced configuration-family selection based on High--High hit rate and validated it on held-out data.
Selection substantially reduced generated mutants (55.6\%) while enriching jointly high-quality mutants, with minimal impact on median IQ and EQ (RQ4).

Future work will expand this analysis in several directions: including post-training operators and regression tasks, developing ranking strategies for different evaluation budgets, and using early quality signals to filter low-value mutants before expensive executions.
Together, these extensions can support broader quality-driven cost reduction across mutation stages, learning tasks, and practical cost constraints.

\section*{Replication Package}

The replication package is available at \url{https://github.com/zaheedahmed/dl-mutant-quality}.

\begin{acks}
We used Claude.ai for language polishing and code assistance.
All scientific and technical content was written and verified by the authors.
\end{acks}

\bibliographystyle{ACM-Reference-Format}
\bibliography{ease2026_refs}


\end{document}